\documentclass[twocolumn,showpacs,aps,prd,superscriptaddress]{revtex4}

\usepackage{graphicx}
\usepackage{dcolumn}
\usepackage{amsmath}
\usepackage{epsfig}

\input pubboard/babarsym
\def\Dspstar{\ensuremath{D_s^{*+}}\xspace}
\def\Dsps{\ensuremath{D_s^{(*)+}}\xspace}
\def\Dsphipi{\ensuremath{\Ds \rightarrow \phi \pip}\xspace}
\def\Dsgamma{\ensuremath{\Dspstar \rightarrow \Ds \gamma}\xspace}

\newcommand{\BABARPubYear}    {01}
\newcommand{\BABARPubNumber}  {17}
\newcommand{\SLACPubNumber} {9131}

\def\figurebox#1#2#3{%
    \def\arg{#3}%
     \ifx\arg\empty
    {\hfill\vbox{\hsize#2\hrule\hbox to #2{\vrule\hfill\vbox to #1{\hsize#2\vfill}\vrule}\hrule}\hfill}%
    \else
    {\hfill\epsfbox{#3}\hfill}%
    \fi}

\long\def\inst#1{\par\nobreak\kern 4pt\nobreak
    {\it #1}\par\vskip 10pt plus 3pt minus 3pt}

\begin{document}

\preprint{\babar-PUB-\BABARPubYear/\BABARPubNumber} 
\preprint{SLAC-PUB-\SLACPubNumber} 

\begin{flushleft}
\babar-PUB-\BABARPubYear/\BABARPubNumber\\
SLAC-PUB-\SLACPubNumber\\
%%hep-ex/\HEPEXNumber\\ [10mm]
\end{flushleft}

\title{
{\large \bf \boldmath Measurement of \Ds and \Dspstar Production in
\B Meson Decays and from Continuum \epem Annihilation at 
$\sqrt{s}=10.6\gev$}
\begin{center}
\vskip 5mm
The \babar\ Collaboration
\end{center}
}

%% author list as of 05-Sep-2001 (584 authors)
%
\author{B.~Aubert}
\author{D.~Boutigny}
\author{J.-M.~Gaillard}
\author{A.~Hicheur}
\author{Y.~Karyotakis}
\author{J.~P.~Lees}
\author{P.~Robbe}
\author{V.~Tisserand}
\affiliation{Laboratoire de Physique des Particules, F-74941 Annecy-le-Vieux, France }
\author{A.~Palano}
\author{A.~Pompili}
\affiliation{Universit\`a di Bari, Dipartimento di Fisica and INFN, I-70126 Bari, Italy }
% ...
\collaboration{The \babar\ Collaboration}
\noaffiliation
 
\date{\today}

\begin{abstract}

New measurements of \Ds and \Dspstar meson production rates 
from \B decays and from \qqbar continuum events near the \FourS resonance
are presented. Using 20.8\invfb 
of data on the \FourS resonance and 2.6\invfb off resonance,
we find the inclusive branching fractions
$\BR(B\rightarrow \Ds X) = (10.93\pm0.19\pm0.58\pm2.73)\%$ and 
$\BR(B\rightarrow \Dspstar X) = (7.9\pm0.8\pm0.7\pm2.0)\%$,
where the first error is statistical, the second is systematic,
and the third is due to the \Dsphipi branching fraction uncertainty.
The branching fractions 
$\Sigma\BR(B\rightarrow \Dsps \Dbar^{(*)}) = (5.07\pm0.14\pm0.30\pm1.27)\%$
and 
$\Sigma\BR(B\rightarrow \Dspstar \Dbar^{(*)}) = (4.1\pm0.2\pm0.4\pm1.0)\%$
are determined from 
the \Dsps momentum spectra.

\end{abstract}

\maketitle

\section{INTRODUCTION}

\begin{figure}
\begin{center}
\includegraphics[width=0.75\linewidth]{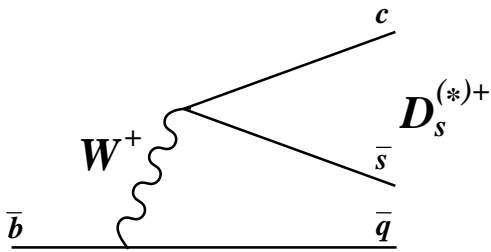}
\caption{
The main spectator diagram leading to the production of \Dsps mesons
in \B decays.}
\label{fig:diagram} 
\end{center}
\end{figure}

The decay of \B mesons into final states involving a \Dsps
provides an opportunity to study the production mechanisms for
$c\overline{s}$ quark pairs$^1$.
\footnotetext[1]{Reference in this paper to a specific decay 
channel or state
also implies the charge-conjugate decay or state. The notation \Dsps means
either \Ds or \Dspstar. $B\rightarrow \Dsps \Dbar^{(*)}$ 
is a general representation for any of the
modes with $c\overline{s}$ and  $\overline{c}q$ states including their excited states.
The notation $B\to\Dsps X$ also implies $\overline{B}\to\Dsps X$.}
Although several diagrams can lead to \Dsps production in \B decays,
the dominant source~\cite{cleo:ssbar} is expected to be external $W^+\to c\overline{s}$
emission (Fig.~\ref{fig:diagram}). A precise knowledge of this production rate
remains interesting in light of continuing theoretical difficulties~\cite{bigi:1}
in accounting for the measurements of both the semileptonic branching
fraction and the inclusive charm production rate in \B decays. Indeed, it has been
noted that an enhanced \B decay rate to charm 
would help explain the small observed semileptonic rate~\cite{FWD:1}.

It is possible to produce \Dsps mesons in \qqbar  events 
from continuum \epem annihilation.
The process of fragmentation ({\it i.e.,} formation of hadrons) 
is non-perturbative and can only be modeled phenomenologically.
The ratio of vector to pseudoscalar production rates is of 
particular interest for testing such models.
The \Ds system is well suited to measure this quantity because 
the $c\overline{s}$ states with $L=1$
have not been observed to decay to either \Ds or \Dspstar mesons. 

In this report, measurements of $B\rightarrow \Ds X$ 
and $B\rightarrow \Dspstar X$ production rates and momentum spectra
are presented. 
We also determine the production cross section for \Ds and \Dspstar mesons
in continuum events.

\section{THE \babar\ DETECTOR AND DATA SET}

The data used for this analysis
were collected with the \babar\ detector~\cite{babar} 
at the \pep2 asymmetric-energy collider~\cite{pep} at the Stanford 
Linear Accelerator Center.
An integrated luminosity of 20.8\invfb was recorded  in 1999 and 2000 at
the \FourS resonance (``on-resonance'') corresponding
to about 22.7$\times$10$^6$ produced \BB pairs, and 2.6\invfb
at an energy about 40\,MeV below the \FourS mass (``off-resonance''). 
A detailed description of the \babar\ detector 
can be found in Ref.~\cite{babar}. Only the components of the detector
most crucial to this analysis are summarized below.

A five-layer double-sided silicon vertex tracker (SVT) and a 40-layer central
drift chamber (DCH) filled with helium-based gas are used to measure the momenta
of charged particles. The tracking system covers 92\% of the solid angle in
the center-of-mass frame and lies within a 1.5-T solenoidal magnetic field. For
charged-particle identification, ionization-energy loss (\dedx) in the
DCH and SVT, and Cherenkov radiation detected in a ring-imaging 
device (DIRC) are used. Photons are identified and measured by a CsI(Tl) 
electromagnetic calorimeter.

\section{THE \Ds AND \Dspstar SELECTION}

Only the decay mode \Dsphipi with $\phi \rightarrow { K^+ K^-}$
is used since it has the best signal-to-background ratio.
Charged tracks are required to originate within $\pm$10\,cm 
of the interaction point
along the beam direction and $\pm$1.5\,cm in the transverse plane, 
and to leave at least 12 hits in the DCH.

Positive kaon identification is required 
for the tracks forming the candidate $\phi$ meson. This is based on
\dedx information from the DCH and SVT,
and the Cherenkov angle and the number of photons  measured with the DIRC.
The kaon selection is based on the likelihood calculated for
each detector component and uses, for each track, the ratio of
likelihoods for the pion and the kaon mass hypotheses, $L_\pi/L_K$.
If this ratio is less than unity for at least one of the detector
subsystems, the particle is selected as a ``loose'' kaon candidate.
A ``tight'' identification criterion is also used in the analysis, based on
the product of the likelihoods for each detector component.
In this case the track is considered a kaon  if the ratio 
of these product likelihoods for the pion- and kaon-mass hypotheses 
is less than unity.

Three charged tracks originating from a common vertex
are combined to form a \Ds candidate. 
Two oppositely charged tracks must be identified as
kaons with the ``loose'' criterion, and at least one of them must
pass the ``tight'' criterion.
No identification criteria are applied to the pion from \Ds decay.
The reconstructed invariant mass of the ${K^+K^-}$  candidates must be within 8\mevcc
of the nominal $\phi$ mass~\cite{pdg}.
In the decay \Dsphipi, the $\phi$ meson is polarized longitudinally
and therefore the angular distribution of the kaons has a
$\cos^2\theta_{H}$ dependence, where
$\theta_{H}$ is the angle between the $K^+$
and \Ds in the $\phi$ rest frame.
We require
$|\cos\theta_{H}|>0.3$, 
which Monte Carlo studies show retains 
97\% of the signal while rejecting about 30\% of the background.
 
\begin{figure}[tb]
\begin{center}
\includegraphics[width=0.8\linewidth]{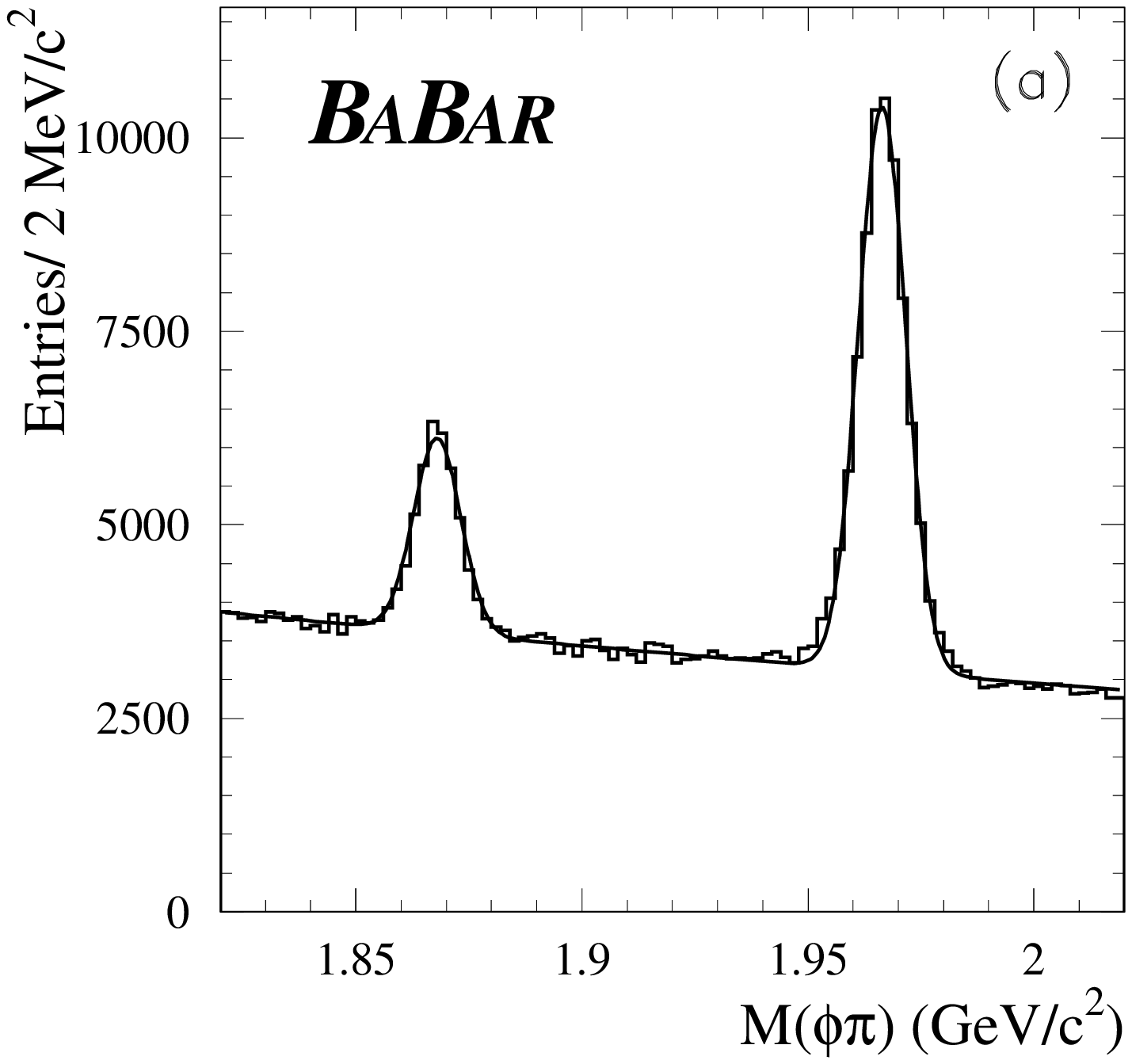}
\includegraphics[width=0.8\linewidth]{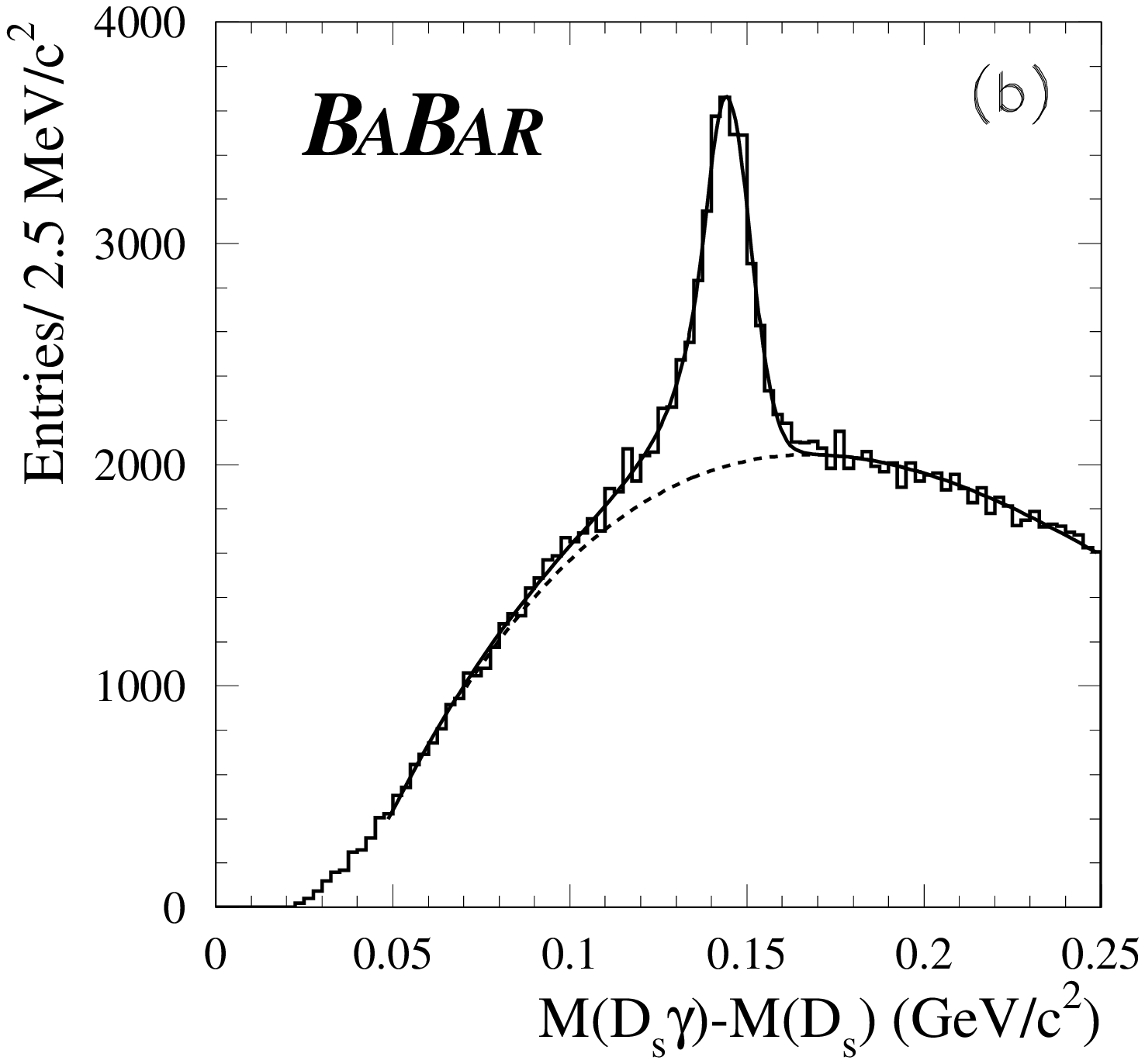}
\caption{
        (a) The $\phi\pi$ invariant mass spectrum. 
	In addition to the \Ds signal, candidates for 
        the Cabibbo-suppressed decay $\Dp \rightarrow \phi \pip$ are also observed.
        The fit function is a single Gaussian for each peak, with widths
        constrained to be equal, plus an exponential background.
	(b) Distribution of the mass difference
        $\Delta M = M_{\Ds\gamma} -  M_{\Ds}$.
        The fit function is a Crystal Ball function
        for the signal plus a threshold function, as described in the text.
}
\label{fig:mass}
\end{center}
\end{figure}

With these requirements, signals for $\Ds\to\phi\pip$
and the Cabibbo-suppressed decay $\Dp \rightarrow \phi \pip$ 
are readily observed (Fig.~\ref{fig:mass}a).
The \Ds and \Dp peaks are both fit with single Gaussian distributions 
with a common free width.
We model the combinatorial background 
with an exponential function. 
From the fit a \Ds signal of $47794 \pm 311$ events is found with
a mass difference
$m(\Ds)-m(\Dp)$ of $98.4\pm0.1\pm0.3\mevcc$. The first error on the latter
is statistical, and the second is systematic, obtained 
from a study of the mass difference as a function of momentum in both data 
and Monte Carlo simulation. Although the uncertainties in the absolute mass scale 
are on the order of several \mevcc, the systematic error in the determination 
of the \Ds and \Dp mass difference is much smaller, 
since many sources of error cancel.

Candidate \Dspstar mesons are reconstructed in the decay \Dsgamma,
with the subsequent decay \Dsphipi.
\Ds candidates are selected by requiring the $\phi\pi$ invariant mass to
be within 2.5 standard deviations ($\sigma$) of the fitted peak value. 
These \Ds candidates are then combined with photon candidates in the event.
Photon candidates are required to satisfy
$E_\gamma > 50$\mev, where $E_{\gamma}$
is the photon energy in the laboratory frame, and
$E_\gamma^* > 110$\mev, where $E^*_{\gamma}$
is the photon energy in the \FourS center-of-mass.
When combined with any other photon in the event,
the photon candidate should not form a $\pi^0$, defined by
a total center-of-mass energy $E_{\gamma\gamma}^* > 200$\mev and
an invariant mass $115 < M_{\gamma\gamma} < 155$\mevcc.
The distribution of the mass difference
$\Delta M = M(\Ds \gamma)-M(\Ds)$
is shown in Fig.~\ref{fig:mass}b.

The $\Delta M$ distribution of the signal  
is parameterized with an asymmetric function to account for
energy leakage and calorimeter shower shape fluctuations.
The signal is modeled with a  
Crystal Ball function~\cite{CBfunc}, which incorporates 
a Gaussian core with a power-law tail toward lower masses.
For the background, a threshold function
\begin{equation*}
f(\Delta M) = p_1 (\Delta M - p_2)^{p_3} e^{p_4(\Delta M-p_2)}
\end{equation*}
is used, where the four parameters $p_i$ are free in the fit.
After ensuring that the connection point between the
Gaussian and power-law tail
does not depend on momentum
and agrees with Monte Carlo simulation, this parameter 
has been fixed to 0.89$\sigma$ in the final fit. 
A signal with $14392\pm 376$ \Dspstar events is observed.

\section{ Extraction of \Dsps momentum spectra}

The momentum spectrum of \Ds mesons in the \epem center-of-mass frame is extracted 
by fitting the $\phi\pi$ invariant mass distribution for 24 ranges of \Ds 
candidate momentum. These ranges are 200\mevc wide,
which is much larger than the momentum resolution ($\approx 6$\mevc).
The same function with two single-Gaussians 
described above for the fit to the full mass distribution
is used as well for the individual momentum bins.
Since there are many more events
in the on-resonance data sample, the number of \Ds in 
the off-resonance data is extracted with
the Gaussian parameters  
($M_{\Dp}$, $M_{\Ds}$ and $\sigma$) fixed to the values obtained from
the on-resonance data.

The center-of-mass momentum spectrum for \Dspstar
mesons is extracted by fitting the $\Delta M$
invariant mass distribution in 250\mevc wide \Dspstar momentum ranges.
We use a larger range because the \Dspstar yield is lower.
The $\Delta M$ distributions are modeled with a  
Crystal Ball function for the signal and a threshold function for
the background
as described above for the fit to the full distribution.
The off-resonance data are again fit
with the Gaussian parameters ($\overline {x}$ and $\sigma$)
fixed to the values obtained from the on-resonance data.

\begin{figure}[tb]
\begin{center}
\includegraphics[width=0.8\linewidth]{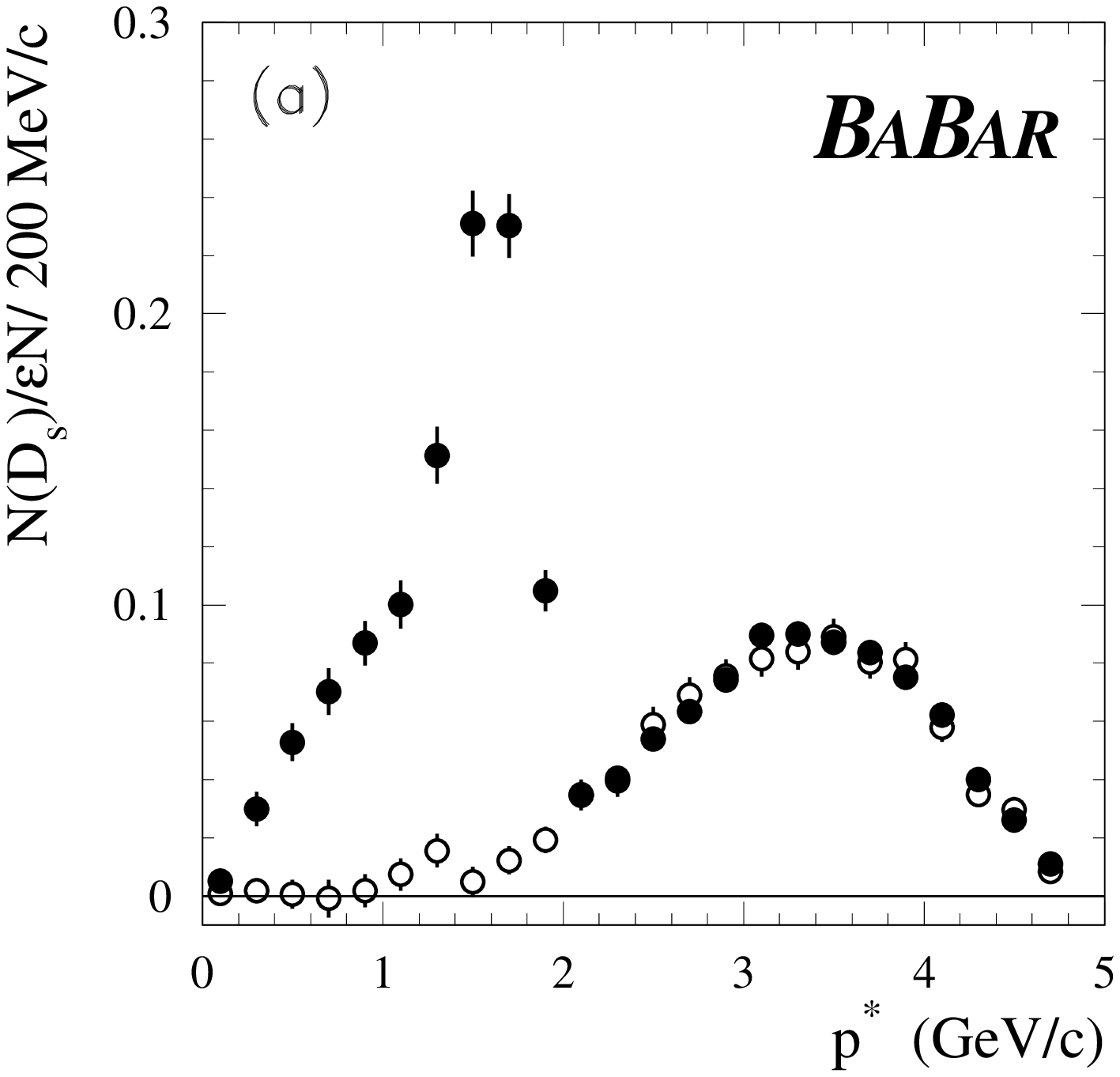}
\includegraphics[width=0.8\linewidth]{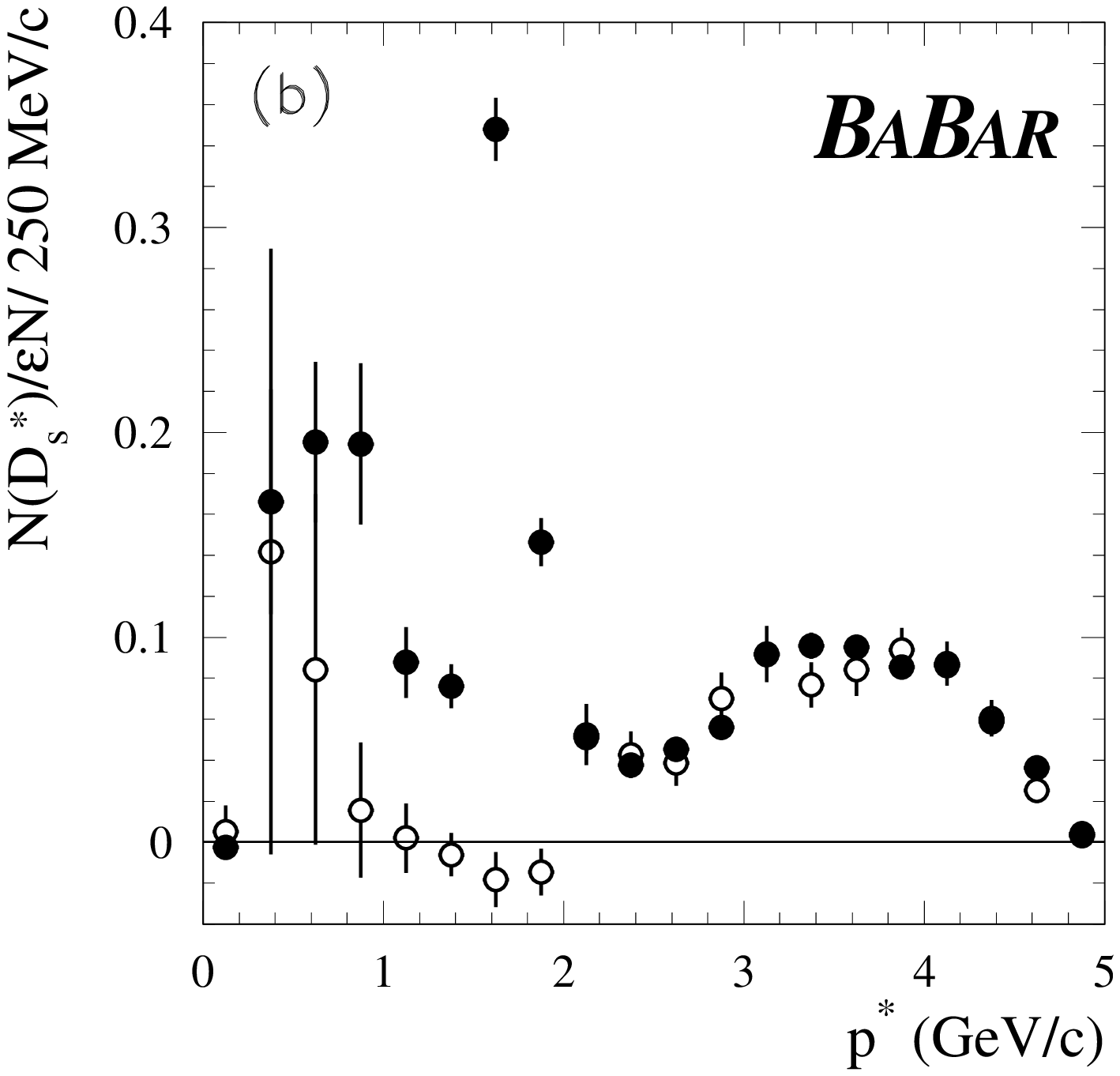}
\caption{
Efficiency-corrected center-of-mass momentum spectra for 
(a) \Ds and (b) \Dspstar for on-resonance (filled circles)
and scaled off-resonance data (open circles).
}
\label{fig:nev}
\end{center}
\end{figure}

The efficiency $\epsilon$, obtained from Monte Carlo simulation of \BB and 
\ccbar events,  varies as a function of the \Dsps center-of-mass momentum $p^*$. 
The efficiency ranges from 20\% (5\%) when the \Ds(\Dspstar) is
at rest to 40\% (20\%) for $p^* = 5 \gevc$.
The efficiency-corrected momentum spectra of \Ds and \Dspstar are shown in  Fig.~\ref{fig:nev}.

\section{INCLUSIVE BRANCHING FRACTIONS}

The \Ds and \Dspstar production cross sections  
in \qqbar continuum are obtained by 
integrating the momentum spectra obtained from
the off-resonance data. 
This gives
$$\sigma (e^+e^-\rightarrow D^{\pm}_sX)\times\BR(\Dsphipi) = 
7.55\pm0.20\pm0.34\ \pb,$$
$$\sigma (e^+e^-\rightarrow D^{*\pm}_sX)\times\BR(\Dsphipi) = 
5.8\pm0.7\pm0.5\ \pb,$$
where the first error is statistical and second systematic.
Sources of systematic error are listed in Table~\ref{tab:syst}.
These include the statistical precision of the Monte Carlo determination
of the efficiency, the luminosity uncertainty, and contributions from 
residual uncertainties on tracking (1.2\% per track) and particle 
identification efficiencies, which are determined from control samples
in data. In addition, for the $D^{*\pm}_sX$ measurement, there
are contributions from the uncertain signal shape, and residual
uncertainties on the photon and
$\piz$ veto efficiencies, again determined with control samples.
 
In order to determine the momentum spectra for \Dsps mesons from \B meson decays,
the off-resonance data are scaled by 
the on- to off-resonance luminosity ratio and then subtracted bin-by-bin from 
the on-resonance data. 
Integrating the resulting spectrum
after continuum subtraction and efficiency correction
gives a total \Ds yield from \B meson 
decays of $87711\pm 1485$ events. This corresponds to 
an inclusive branching fraction of
$$
\BR(B\rightarrow D_s^+X) = \Biggl[(10.93\pm0.19\pm0.58)\times 
\frac{(3.6\pm0.9)\%}{\BR(D_s^+\rightarrow\phi\pi^+)}\Biggr]\%.
$$
Likewise, the total \Dspstar yield from \B meson 
decays is $60047\pm 6201$ events, leading to the 
inclusive branching fraction of
$$\BR(B\rightarrow D_s^{*+}X) = \Biggl[(7.9\pm0.8\pm0.7)\times 
\frac{(3.6\pm0.9)\%}{\BR(D_s^+\rightarrow\phi\pi^+)}\Biggr]\%.$$
In the results above, the first error is statistical and
the second is systematic. The dominant error, due to the uncertainty 
in the \Dsphipi branching fraction of ($3.6\pm 0.9$)\%~\cite{pdg}, 
is shown separately.
It is important to note that, with this method, the result is independent of
any assumption regarding the shape of the fragmentation function.
The various contributions to the systematic error are listed in Table~\ref{tab:syst}.
In addition to the sources already noted above,
the uncertainty in the shape of the background
impacts this measurement, particularly in the lower
momentum bins. 
This contribution is estimated with the use of different 
parameterizations for the background shape and different methods for
handling the continuum subtraction. The efficiency variation 
over the width of the momentum bins is also included
as an additional systematic error.

\begin{table}
\begin{center}
\caption{ Systematic errors for cross section and branching fraction measurements.}
\label{tab:syst}
\begin{tabular}{lcccc}
\hline
\hline
Source  & \multicolumn{4}{c}{Fractional Error (\%)}      \\ \hline
        & \multicolumn{2}{c}{Continuum} & \multicolumn{2}{c}{$B$ Decays} \\ \hline
        & $\Ds X$ & $\Dspstar X$ & $\Ds X$ & $\Dspstar X$ \\ \hline
Signal shape                                 &     & 3.0 & 0.5 & 3.0 \\
Background subtraction                       &     &     & 0.4 & 4.2 \\
Monte Carlo statistics                       & 1.0 & 4.8 & 2.5 & 4.2 \\
Bin width                                    &     &     & 1.4 & 2.0 \\ 
\hline
Total for \Dsps yield                        & 1.0 & 5.7 & 2.9 & 7.0 \\
Luminosity/N(\BB)                            & 1.5 & 1.5 & 1.6 & 1.6 \\
$\BR(\phi\rightarrow { K^+K^-})$             & 1.6 & 1.6 & 1.6 & 1.6 \\
Particle identification                      & 1.0 & 1.0 & 1.0 & 1.0 \\
Tracking efficiency                          & 3.6 & 3.6 & 3.6 & 3.6 \\ 
$\BR(\Dsgamma)$                              &     & 2.7 &     & 2.7 \\
Photon efficiency                            &     & 1.3 &     & 1.3 \\
$\pi^0$ veto                                 &     & 2.7 &     & 2.7 \\ 
\hline
Total systematic error                       & 4.5 & 8.2 & 5.3 & 9.0 \\ 
\hline \hline
\end{tabular}
\end{center}
\end{table}

\section{ FITS TO \Dsps MOMENTUM SPECTRA}

By fitting the \Dsps momentum spectrum, relative branching
fractions of \B decays to different final states containing \Dsps
mesons are obtained.
In the \FourS rest frame, two-body \B decays 
produce \Dsps mesons with a momentum spectrum 
about 300\mevc wide.
In \B decays, the \Dsps momentum spectrum is essentially governed by the 
production of direct \Dsps.
Other $c\overline{s}$ states (with $L=1$), such as $D_{s1}^+(2536)$ and $D_{s2}^{*+}(2573)$,
primarily decay to $D^{(*)}K$. 
Because \Dspstar decays to $\Ds\gamma$ or $\Ds\pi^0$, the \Ds momentum 
distribution is slightly broader and shifted downward
compared to direct production from $B\rightarrow \Ds X$.

\begin{figure}[tb]
\begin{center}
\includegraphics[width=0.8\linewidth]{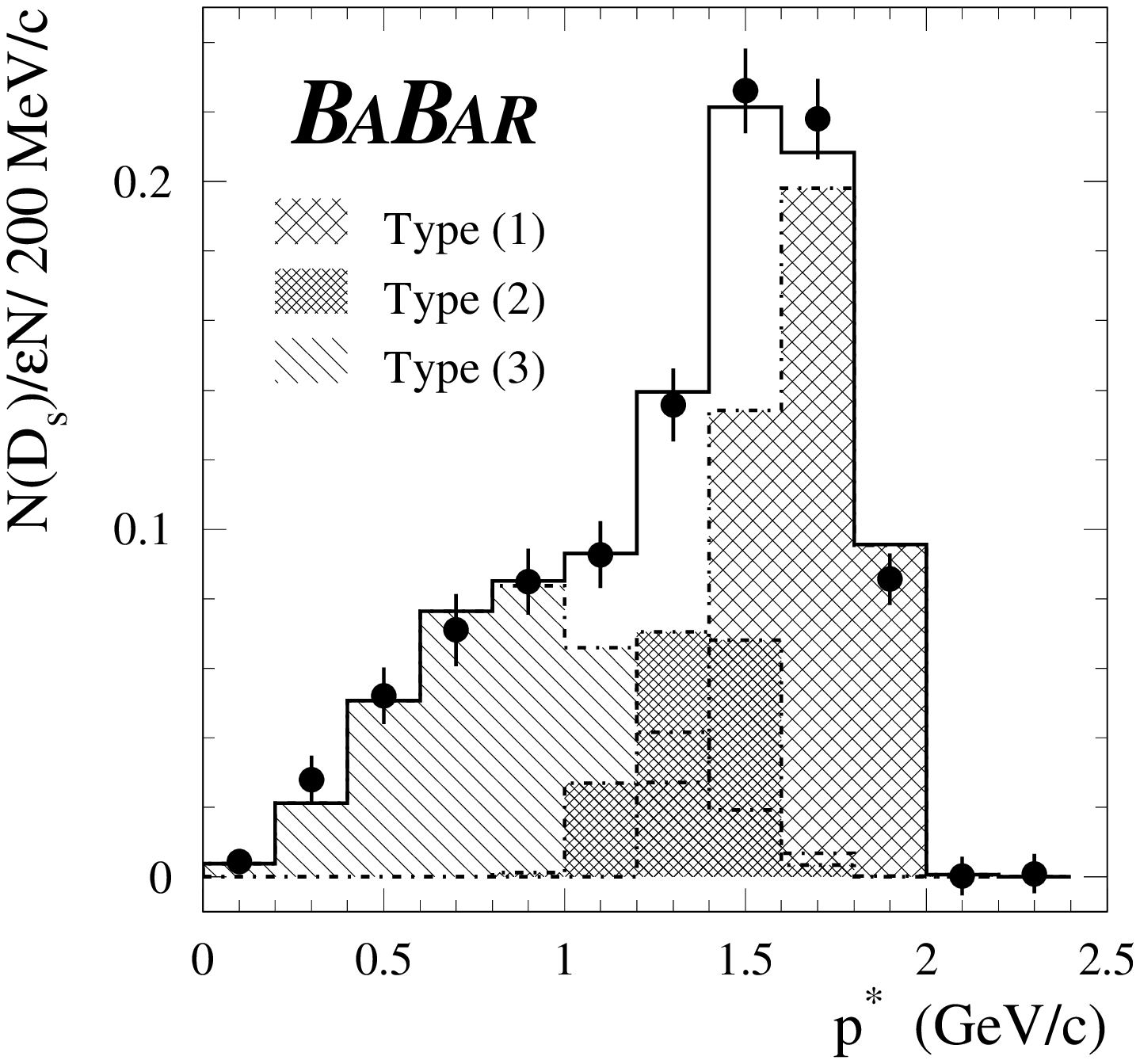}
\includegraphics[width=0.8\linewidth]{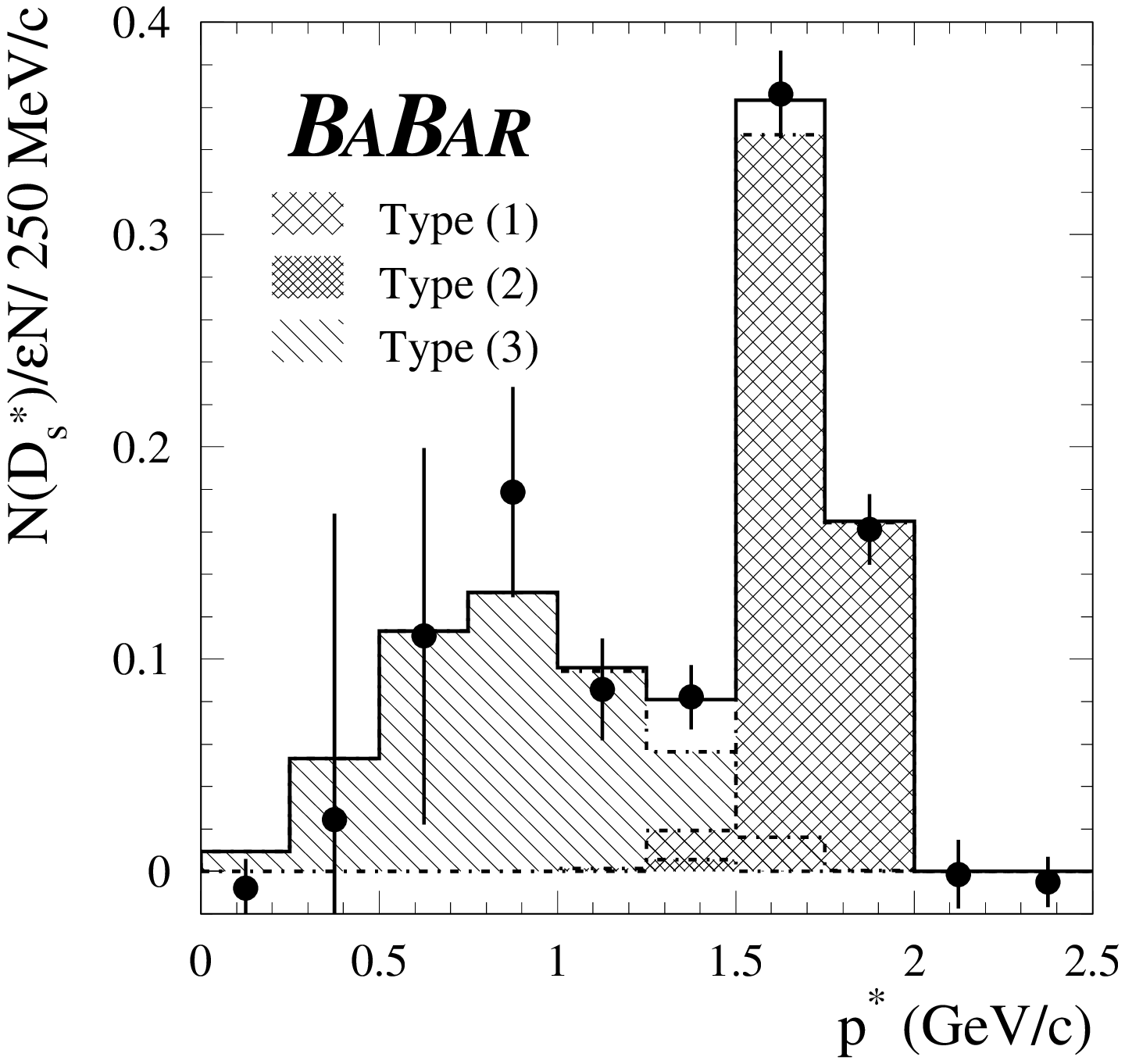}
\caption{
Fit results for (a) \Ds and (b) \Dspstar momentum spectra. 
The data are dots with error bars, and the histograms are the components 
of the fit function described in the text. Type (1) is
$B\rightarrow \Dsps \Dbar^{(*)}$, Type (2) is
$B\rightarrow \Dsps \Dbar^{**}$ and Type (3) is
$B\rightarrow \Dsps \Dbar^{(*)} \pi/\rho/\omega$.
The solid histogram is the sum of the three components.
}
\label{fig:fit}
\end{center}
\end{figure}

Three different sources of \Dsps mesons in \B decays are considered for 
the fits to the momentum spectra.
\begin{enumerate}
\renewcommand{\labelenumi}
                {(\arabic{enumi})}
\item $B \rightarrow \Dsps \Dbar^{(*)}$ decays.
The relative branching fractions of the individual channels
can be taken either from existing measurements~\cite{cleo:dsinc}
or from predictions that assume factorization~\cite{th_BSW,th_rosner,th_neubert}.
The fit is performed for both cases, with the assumption
$f_{\Dspstar}=f_{\Ds}$ for the theoretical models, where   
$f_{\Dsps}$ are the \Dsps decay constants.
\item $B\rightarrow \Dsps \Dbar^{**}$ decays. 
Four $\Dbar^{**}$ states are considered:
$\Dbar_0^*(j=1/2)$, $\Dbar_1(2420)$, $\Dbar_1(j=1/2)$ and $\Dbar_2^*(2460)$.
Observation of $B\rightarrow \Dsps \Dbar^{**}$ decays
was recently reported by CLEO~\cite{cleo:preco}.

\item Three-body ${ B\rightarrow \Dsps \Dbar^{(*)} \pi/\rho/\omega }$ decays.
Since little is known about these decays, 
they are attributed equal weights, and the momentum 
distributions are generated according
to phase space.
\end{enumerate}

Minimum-$\chi^2$ fits to the \Dsps\ momentum spectra are performed,
where the total number of \Dsps events and the fractions
of the source (1) and (2) contributions are free parameters.
From the fits to the \Ds\ and \Dspstar\ spectra, 
the ratios of two-body modes [source (1)] to the total inclusive rate are
determined to be
\begin{eqnarray}
&&\frac{\Sigma\BR(B\rightarrow \Dsps \Dbar^{(*)})} 
{\BR(B\rightarrow \Ds X)} =
(46.4 \pm 1.3 \pm 1.4 \pm 0.6) \%, \nonumber\\
&&\frac{\Sigma\BR(B\rightarrow \Dspstar \Dbar^{(*)})} 
{\BR(B\rightarrow \Dspstar X) } =
(53.3 \pm 3.7 \pm 3.1 \pm 2.1) \%.\nonumber
\end{eqnarray}
The first error is statistical. The second error represents the systematic 
error due to the limited Monte Carlo statistics and the background parameterization.

The last error is due to the model uncertainty. 
It is obtained by varying the relative 
fractions of the modes contributing to each source of \Dsps listed above.
The fit is performed with alternative assumptions for the relative 
contributions of the modes in source (1) 
taken from theoretical predictions and measurements. 
Different weights for
${ B\rightarrow D_s^+ \Dbar^{**}}$ 
and ${ B\rightarrow D_s^{*+} \Dbar^{**}},$
as well as different relative branching fractions 
of the four modes within source (2), are used.
For source (3),
either $B\rightarrow D_s^{(*)} \Dbar^{(*)} \pi$ or 
$B\rightarrow D_s^{(*)} \Dbar^{(*)} \rho/\omega$ 
is assumed to be dominant. The $\chi^2$ of the fit for the
inclusive \Dspstar momentum spectrum is lowest when the contribution of
$B\rightarrow D_s^{(*)} \Dbar^{(*)} \rho/\omega$ 
is dominant compared to
$B\rightarrow D_s^{(*)} \Dbar^{(*)} \pi$.
The results of the fits to the \Dsps momentum spectra are shown in 
Fig.~\ref{fig:fit} under the assumption of equal weights for 
the individual contributions within sources (2) and (3), 
and with the weights of the individual modes of source (1) 
taken from~\cite{th_neubert}.

The sum of branching fractions for the two-body 
$B\rightarrow D_s^{(*)} \Dbar^{(*)}$ decays
are obtained from the fits to the \Dsps momentum spectra, 
where the yield from each source is a free parameter.
We find
\begin{eqnarray}
\Sigma\BR(B\rightarrow \Dsps \Dbar^{(*)}) &=& 
(5.07\pm0.14\pm0.30\pm1.27)\%,\nonumber\\
\Sigma\BR(B\rightarrow D_s^{*+} \Dbar^{(*)}) &=& 
(4.1\pm0.2\pm0.4\pm1.0)\%,\nonumber
\end{eqnarray}
where the first error is statistical, the second is systematic, 
and the third is due to the \Dsphipi branching fraction uncertainty.
The systematic error includes contributions from
the $B\to\Dsps X$ branching fractions, 
the relative contributions of source (1),
and model dependence of the source spectra.
The sum of the two-body modes is reasonably separated in the momentum spectra
from the other components.  Therefore, the fractional error on the sum of the
two-body modes is smaller than the fractional error on the $B\to\Dsps X$ branching
fraction or the relative two-body branching ratio.

%==================================================================
%
%       SUMMARY
%
%==================================================================
\section{SUMMARY}

In summary, the branching fractions for inclusive 
$B\rightarrow\Dsps X$ production have been determined as well as
the \Dsps production cross sections from continuum events
at center-of-mass energies about 40\mev below the \FourS mass.
Our more precise results for the \Ds are in agreement with previous
measurements~\cite{cleo:dsinc,argus}, while the \Dspstar measurements
are new. 
In contrast to previous results,
our measurements do not rely on any assumptions regarding 
the shape of the fragmentation function. 
Finally, fits to the \Dsps momentum spectra provide
relative yields and branching fractions for two-body 
$B\rightarrow \Dsps \Dbar^{(*)}$ and 
$B\rightarrow D_s^{*+} \Dbar^{(*)}$ decays.
The mass difference $m(\Ds)-m(\Dp)$ has also been measured.

%==================================================================
%
%       ACKNOWLEDGMENTS
%
%==================================================================
\section{ACKNOWLEDGMENTS}
We are grateful for the excellent luminosity and machine conditions
provided by our \pep2\ colleagues.
The collaborating institutions wish to thank 
SLAC for its support and kind hospitality. 
This work is supported by
DOE
and NSF (USA),
NSERC (Canada),
IHEP (China),
CEA and
CNRS-IN2P3
(France),
BMBF
(Germany),
INFN (Italy),
NFR (Norway),
MIST (Russia), and
PPARC (United Kingdom). 
Individuals have received support from the Swiss NSF, 
A.~P.~Sloan Foundation, 
Research Corporation,
and Alexander von Humboldt Foundation.

%%%%%%%%%%%%%%%%%%%%%%%%%%%%%%%%%%%%%%%%%%%%%%%%%%%%%%%%%%%%%%%%%%%


\begin{thebibliography}{99}
\bibitem{CBfunc}
The $\Delta M$ distribution for the $\Ds\gamma$ signal is fit
with the Crystal Ball function
  \[
  f(x) = N \cdot \left\{
  \begin{array}{ll}
    \exp (-\frac{(x - \overline {x})^2}{2 \; \sigma^2})  &;  
( \overline {x} - x)/\sigma >   \alpha \\
    A \times (B - \frac{x - \overline {x}}{\sigma})^{-n} &;  
( \overline {x} - x)/\sigma \leq \alpha \\
  \end{array} \right. \]
  where $A \equiv \left ( \frac{n}{|\alpha|} \right ) ^n 
\times \exp(-|\alpha|^2/2)$ and
  $B \equiv \frac{n}{|\alpha|} - |\alpha|$.
  $N$ is a normalization factor, $\overline x$ and $\sigma$ are the peak position and width
  of the Gaussian portion of the function, $\alpha$ is the point at which the 
function changes to the power function and $n$ is the exponent of the power function.
  $A$ and $B$ are defined so that the function 
and its first derivative are continuous at $\alpha$.
  More details can be found in D.\ Antreasyan, Crystal Ball Note 321 (1983).
\end{thebibliography}
\end{document}